\documentclass[11pt]{article}
\usepackage[a4paper, left=1in, right=1in, top=1in, bottom=1in]{geometry}
\usepackage{graphicx,url} % Required for inserting images
\usepackage{authblk,longtable,tabularx,rotating,multirow,url}
\usepackage{amsmath,natbib,comment,subcaption}
\usepackage{amsfonts,color,gensymb}
\usepackage{amssymb}
\usepackage{verbatim}
\usepackage{draftwatermark}
\SetWatermarkText{Confidential}
\SetWatermarkScale{5}
\SetWatermarkLightness{0.85}

\title{Impact of rainfall risk on rice production: realized volatility in mean model}

\author[1]{Soham Ghosh}%\thanks{Corresponding author: phd2001161004@iiti.ac.in}}
\author[2]{Sujay Mukhoti}
\author[1]{Pritee Sharma}

\affil[1]{
    {Department of Economics, School of Humanities and Social Sciences},
    {Indian Institute of Technology Indore}, 
    {Indore},
    {453552}, 
    {Madhya Pradesh},
    {India}}
            
\affil[2]{
    Operations Management and Quantitative Techniques Area, Indian Institute of Management Indore,Indore, 453556, Madhya Pradesh, India}

\date{February 2025}

\begin{document}

\maketitle

\begin{abstract}
Rural economies are largely dependent upon agriculture, which is greatly determined by climatic conditions such as rainfall. This study aims to forecast agricultural production in Maharashtra, India, which utilises annual data from the year 1962 to 2021. Since rainfall plays a major role with respect to the crop yield, we analyze the impact of rainfall on crop yield using four time series models that includes ARIMA, ARIMAX, GARCH-ARIMA and GARCH-ARIMAX. We take advantage of rainfall as an external regressor to examine if it contributes to the performance of the model. 1-step, 2-step, and 3-step ahead forecasts are obtained and the model performance is assessed using MAE and RMSE. The models are able to more accurately predict when using rainfall as a predictor compared to when solely dependant on historical production trends (more improved outcomes are seen in the ARIMAX and GARCH-ARIMAX models). As such, these findings underscore the need for climate-aware forecasting techniques that provide useful information to policymakers and farmers to aid in agricultural planning.
\end{abstract}

\section{Introduction}

Rice is a major food for a large portion of the global population and it plays a crucial role in ensuring food security and economic stability in India \citep{maraseni2018international}. The country’s agricultural sector is heavily reliant on rice production, which is influenced by several meteorological, economic, and policy-related factors \citep{green2016dietary}. In recent years, India has emerged as one of the world’s largest producers and consumers of rice \citep{mohan2010can}. During the 2023–24 agricultural year, the country produced an estimated 137.83 million metric tons of rice, accounting for about $26\%$ of the global production \citep{glauber2024global}. Accurate forecasting of rice production is essential for policymakers, farmers, and other stakeholders to make well-informed decisions on agricultural planning, food supply management, and price stabilization. In this work, we analyse how rainfall variability is influencing rice production in an Indian state, Maharastra, where agricultural productivity is vulnerable to climatic fluctuations.

Maharashtra is an important India state in terms of agricultural production because of its evolving agroclimatic environment  \citep{thaware2014status}. The evolving environment ensures production of a wide range of crops, including rice. Punjab, West Bengal, and Uttar Pradesh are well known for producing large quantity of rice \citep{mahajan2017rice}. In contrast to these conventionally recognized states, Maharashtra is a special example because of its particular agricultural circumstances. Rice production in Maharastra is extremely sensitive to monsoon rainfall, as a major part of the agricultural land is rainfed. The state experiences major climatic variations, ranging from heavy rainfall in the Konkan region to semi-arid conditions in central Maharashtra \citep{gharke2013brief}. Maharashtra’s rice production fluctuates more due to inconsistent rainfall patterns, making it an important case for studying the impact of precipitation on production. 

Rice is a staple food for a significant section of Maharashtra's population, especially in rural areas of the state \citep{raje2015maternal}. For smallhold farmers, rice remains a primary source of income for their livelihood \citep{shende2015food}. Establishing stability in rice production is essential from both social and economic perspectives. An increasing number of extreme weather events including droughts, unseasonal rains, and heatwaves are affecting Maharastra during recent times. Rice production in Maharastra gets seriously threatened by these climatic events, and forecasting models incorporating climate variability into account are required to improve resilience and adaptation plans. State government frequently carries out strategies associated with crop insurance, irrigation schemes, and minimum support prices (MSP). In this paper, our approach aims to provide policymakers with reliable predictions that can support better decision-making for agricultural planning.

Realized variance serves as a measure of risk and it is defined by quantifying the extent of variability or uncertainty in a time series over a given period \citep{hansen2006realized}. Realized variance requires periodic data to construct an improved representation of volatility compared to traditional variance \citep{barndorff2008designing}. This statistic is extensively applied in the financial econometrics literature to model asset price fluctuations, but it is equally valuable in forecasting climate impacts on agricultural production \citep{degiannakis2022forecasting,bonato2023nino}. In the literature of time series forecasting, realized variance is included as an exogenous variable allowing models to capture fluctuations in external risk factors. In this study, we construct realized variance measures from monthly rainfall data to quantify climate-driven uncertainty in rice production. 

In this work, we introduce four exogenous variables to quantify several aspects of variability in the rainfall series. First one represents the square root of the annual sum of squared rainfall highlighting the extreme precipitation events. Second quantity is defined as the standard deviation of the annual rainfall series capturing the irregularity in rainfall distribution within a year. Third statistic is computed by is computed by summing the squared changes in the log-differenced monthly rainfall and taking the square root, and it captures the overall fluctuation in rainfall time-series throughout the year. Fourth measure represents the dispersion of the log-differenced monthly rainfall around its mean within a year, explaining the intra-annual variability. By incorporating these measures into AutoRegressive Integrated Moving Average with Exogenous Variables (ARIMAX) and Generalized Autoregressive Conditional Heteroskedasticity with AutoRegressive Integrated Moving Average with Exogenous Variables (GARCH-ARIMAX) models, we assess their role in predicting not only the mean production levels but also the time-varying volatility in agricultural output. Additionally, AutoRegressive Integrated Moving Average (ARIMA) and Generalized Autoregressive Conditional Heteroskedasticity with AutoRegressive Integrated Moving Average (GARCH-ARIMA) models are also employed for prediction purposes, serving as benchmarks to measure the effectiveness of including exogenous variables. This modeling framework improves forecast accuracy and provides valuable insights for agricultural planning.

To compare the performance of our models, we deploy both in-sample and out-of-sample validation metrics. We employ the Akaike Information Criterion (AIC) for in-sample model selection \citep{akaike1973second}. The statistic model complexity and goodness of fit by penalizing surplus parameters while preferring models with smaller residual variance. Small values of AIC indicate a more optimal trade-off between accuracy and parsimony. Mean Absolute Error (MAE) is selected for out-of-sample assessment criterion. This statistic quantifies the average magnitude of forecasting errors without considering the direction. MAE provides a measure of forecasting accuracy, ensuring that models with lower values of MAE are preferred. By considering both AIC and MAE, our focus is on identifying the models that demonstrate a better fit to the historically available data and depict robust generalization to unknown future observations.

The remainder of this paper is structured as follows: Section 2 provides an overview of the data and methodology. Section 3 describes the time series models employed for forecasting. Section 4 presents the results and analysis. Finally, Section 5 discusses the key findings and their implications for agricultural policy and planning.

\section{Production forecasting model}

\subsection{Realized variance: a rainfall risk measurement}

Realized variance (RV) is defined as the sum of squared of intraday returns over a given period (weekly, monthly, yearly), and is extensively applied as a measure of volatility \citep{andersen1998answering}. In financial time series literature, an intraday return refers to the price movement of an asset over short time intervals (every minute or hour) in a single day. If $P_t$ represents the price of an asset an time $t$, then the intraday return over the interval $(\Delta t)$ is computed as the change in price in that interval, \[r_t=\log\left(\frac{P_t}{P_{t-\Delta t}}\right)\]. Then, the realized variance is expressed as \[RV(t) = \sum_{i=1}^{t} r_i^2,\] n is the number of observations within the specified day. Adapting the same concept to hydrological time series, we define the realized variation of rainfall as the sum of squared monthly rainfall values within a given year. Specifically, for year t, the realized variation is computed as: \[RV_{rainfall}(t) = \sum_{m=1}^{12} \left\{\log\left(\frac{R_{t,m}}{R_{t,m-1}}\right)\right\}^2\], $R_{t,m}$ denotes the recorded rainfall in month m of year t. Realized variance is a random quantity continuously evolving over time as compared to traditional variance. Realized volatility is computed as the square root of the realized variance. 

Another important property of the realized variance is that it is a consistent estimator of the quadratic variation present in a process under the assumption of absence of white noise and autocorrelation \citep{barndorff2002econometric}. This measure is also applied to estimate the integrated variance. If the log-rainfall $(r_t)$ follows a stochastic process given by the integral, \[r_t = r_0 + \int_0^t \sigma_s^2 dB_s.\] Here, $B_s$ and $\sigma_s$ present the standard Brownian motion and volatility of the actual process respectively. Integrated volatility over the continuous time-interval $[0,t]$ is expressed as, \[\sigma^{2*} = \int_0^t \sigma_s^2 ds.\] \citet{barndorff2002econometric} demonstrates that for large values of $n$, realized variance is a consistent estimator of $\sigma^{2*}$. Additionally, the asymptotic distribution of the realized variance is standard normal \[\sqrt{n}\times\left(\frac{RV(t)-\sigma^{2*}}{\sqrt{2t \int_0^t \sigma_s^4 ds}}\right)\rightarrow N(0,1)\]. To develop an improved understanding about the impact of external variables on rice productivity, we introduce four rainfall risk measures into our econometric models. In comparison to raw precipitation data, which may not capture the risk on agricultural productivity, these four annualized measures based on monthly rainfall data are computed from the concept of realized volatility and serve as a quantitative representation of rainfall risk. 

\subsubsection{Different forms of rainfall risk}

\textbf{Square Root of Annual Sum of Squared Rainfall}:  \\
This measure represents the intensity of extreme precipitation events within a given year and is computed as:
\begin{equation}
     RV_1(t) = \sqrt{\sum_{m=1}^{12} R_{t,m}^2}.
\end{equation}
This transformation scales up higher rainfall values. This allows for a better understanding of non-linear impacts on rice production, where both low and excessive rainfall pose significant risks to agricultural output. \\

\textbf{Annual Rainfall Standard Deviation}:\\ 
This variable is computed as the square root of the variance of monthly rainfall values within each year:
\begin{equation}
    RV_2(t) = \sqrt{\frac{1}{12} \sum_{m=1}^{12} (R_{t,m} - \bar{R_t})^2}
\end{equation}
where \( \bar{R_t} \) is the mean monthly rainfall in year \( t \). This transformation demonstrates the variability of rainfall distribution within a given year. High variability in the rainfall series depicts extreme weather situations, such as irregular patterns in monsoon rainfall, which has an adverse impact on rice production. \\
\textbf{Realized Volatility}\\
This measure is identical to $RV_{rainfall}(t)$ described in the previous section and it depicts sudden changes in rainfall series affecting different stages of crop growth.\\

\textbf{Standard Deviation of Log Differences}\\
This statistic quantifies the dispersion of log differences around their mean within a given year:
\begin{equation}
    RV_4(t) = \frac{1}{12} \sum_{m=1}^{12} \left\{\log\left(\frac{R_{t,m}}{R_{t,m-1}}\right) - \overline{\Delta \log R_t}\right\}^2
\end{equation}
where \( \overline{\Delta \log R_t} \) is the mean of log difference in year \( t \). This measure captures short-term volatility in rainfall patterns, which is particularly relevant for modeling agricultural risk.\\

Agricultural yield particularly in the rainfed regions is heavily influenced by rainfall. Incorporating these risk measures in different time series models, we quantify how different aspects of rainfall variability are affecting rice production dynamics.

\subsection{Time-series models for production forecasting}
We employ four distinct time series models: ARIMA, ARIMAX, GARCH-ARIMA, and GARCH-ARIMAX, to investigate both predictive power and impact of rainfall variability on agricultural productivity in Maharashtra.

\subsubsection{ARIMA Model}

ARIMA is one of the most popular and widely used time series models for forecasting univariate time series data. The model was proposed by \citet{box1970time}. This model has three components: autoregression (AR), differencing (I), and moving average (MA) and is expressed as ARIMA(p, d, q). Here,

\begin{itemize}
    \item[] p represents the number of lagged observations in the autoregressive (AR) component.
    \item[] d denotes the number of times the series needs to be differenced to achieve stationarity.
    \item[] q signifies the number of lagged forecast errors in the moving average (MA) component.
\end{itemize}

The model is mathematically represented as:

\begin{equation}
    Y_t = \alpha + \sum_{i=1}^{p} \phi_i Y_{t-i} + \sum_{j=1}^{q} \theta_j \epsilon_{t-j} + \epsilon_t
\end{equation}

Where:
\begin{itemize}
    \item[] \(Y_t\) is the time series of rice production at time \(t\),
    \item[] \(\alpha\) is a constant term,
    \item[] \(\phi_i\) are the autoregressive parameters for lags \(1\) through \(p\),
    \item[] \(\theta_j\) are the moving average parameters for lags \(1\) through \(q\),
    \item[] \(\epsilon_t\) is the error term.
\end{itemize}

\subsubsection{ARIMAX Model}

The ARIMAX model is an extension of the ARIMA model that incorporates exogenous variables as predictors \citep{pandit2023hybrid}. The approach of integrating external variables may improve the prediction accuracy of the dependent variable. The model is expressed as:

\begin{equation}
    Y_t = \alpha + \sum_{i=1}^{p} \phi_i Y_{t-i} + \sum_{j=1}^{q} \theta_j \epsilon_{t-j} + \sum_{k=1}^{r} \gamma_k X_{t-k} + \epsilon_t
\end{equation}

Here,
\begin{itemize}
    \item[] \(X_{t-k}\) represents the exogenous variables at lag \(k\),
    \item[] \(\gamma_k\) are the coefficients for the exogenous variables at lags \(1\) through \(r\),
\end{itemize}

\subsubsection{GARCH Model Variants}

Traditional time series models are applied under the assumption of constant variance. In real world scenarios, such assumption may not be always true. \citet{engle1982autoregressive} introduced the Autoregressive conditional heteroskedasticity (ARCH) model relaxing the assumption of constant variance. In ARCH models, conditional variance (or volatility) changes over time and is expressed as a function of past errors and unconditional variance remain constant. Generalised ARCH (GARCH) model is an enhanced version of the ARCH models. In this study, four different GARCH models are employed to analyze how different types of volatility in rainfall series are affecting rice production: symmetric GARCH (sGARCH), exponential GARCH (eGARCH), threshold GARCH (gjrGARCH), and integrated GARCH (iGARCH). 

\textbf{Standard GARCH (sGARCH)}

\citet{bollerslev1986generalized} proposed the standard GARCH (sGARCH) model, which allows for a more flexible lag structure. This model represents conditional variance as a function of previous conditional variances and squared errors. The model is expressed as:

\begin{equation}
    \sigma_t^2 = \omega + \sum_{i=1}^{p} \alpha_i \epsilon_{t-i}^2 + \sum_{j=1}^{q} \beta_j \sigma_{t-j}^2
\end{equation}

Here,
\begin{itemize}
    \item[] \(\sigma_t^2\) is the conditional variance at time \(t\),
    \item[] \(\omega\) is a constant term,
    \item[] \(\alpha_i\) are the coefficients for past squared error terms (\(\epsilon_{t-i}^2\)), which represents the short-term shock effect,
    \item[] \(\beta_j\) are the coefficients for past conditional variances (\(\sigma_{t-j}^2\)), capturing the long-term persistence of volatility.
\end{itemize}

Volatility shock refers to sudden or abrupt changes in the volatility of a time series. In the context of rainfall time series, a positive volatility shock indicates changes from stable weather conditions to highly erratic conditions. Similarly, a negative shock suggests moderate rainfall is observed when the series reflects high volatility. In this model, impacts of the positive and negative shocks of the same magnitude on volatility remain identical. 

\textbf{Exponential GARCH (eGARCH)}

Asymmetric effect refers to the situations when positive and negative shocks have significantly different impacts on the future volatility. To incorporate the asymmetric effect in the model, \citet{nelson1991conditional} extends the GARCH modelling framework by presenting the exponential GARCH (eGARCH) model. This model explains the asymmetric effect by modeling the logarithm of the conditional variance as a function of past errors and variances in the following manner:

\begin{equation}
    \ln(\sigma_t^2) = \omega + \sum_{i=1}^{p} \alpha_i \left( \frac{\epsilon_{t-i}}{\sigma_{t-i}} \right) + \sum_{j=1}^{q} \beta_j \ln(\sigma_{t-j}^2) + \sum_{i=1}^{p} \gamma_i \left( \left| \frac{\epsilon_{t-i}}{\sigma_{t-i}} \right| - E \left| \frac{\epsilon_{t-i}}{\sigma_{t-i}} \right| \right)
\end{equation}

Here, \(\gamma_i\) presents the asymmetric response to volatility shocks. \(\gamma_i > 0\) indicates that negative shocks increase future volatility more than positive shocks. This model is useful for explaining leverage effects in agricultural time series.

\textbf{Threshold GARCH (gjrGARCH)}

The threshold GARCH (gjrGARCH) model is presented by \citet{glosten1993relation} and the model also captures the asymmetric effect. In order to consider asymmetric volatility responses, an indicator function is introduced to the sGARCH model, allowing negative shocks to have larger impact on future volatility series compared to positive ones:

\begin{equation}
    \sigma_t^2 = \omega + \sum_{i=1}^{p} \alpha_i \epsilon_{t-i}^2 + \sum_{j=1}^{q} \beta_j \sigma_{t-j}^2 + \sum_{i=1}^{p} \gamma_i I_{t-i} \epsilon_{t-i}^2
\end{equation}

Here, \(I_{t-i}\) is an indicator function that takes the value 1 if \(\epsilon_{t-i} < 0\) (negative shock) and 0 otherwise. The term \(\gamma_i\) measures the additional effect of negative shocks on volatility.

\textbf{Integrated GARCH (iGARCH)}

The integrated GARCH (iGARCH) model is an improved version of the sGARCH model, where the effect of volatility shocks is long-lasting  \citet{engle1986modelling}, suggest the previous shocks are influencing future volatility indefinitely:

\begin{equation}
    \sigma_t^2 = \omega + \sum_{i=1}^{p} \alpha_i \epsilon_{t-i}^2 + \sum_{j=1}^{q} \beta_j \sigma_{t-j}^2
\end{equation}

subject to the constraint:

\begin{equation}
    \sum_{i=1}^{p} \alpha_i + \sum_{j=1}^{q} \beta_j = 1.
\end{equation}

This transformation ensures volatility shocks remain forever. This model is useful for those time series data where volatility clustering demonstrates long-term consequences.

\subsubsection{GARCH in Mean (GARCH-M) Model}

The GARCH-in-Mean (GARCH-M) model extends the conventional GARCH framework. In these models, volatility directly influences the mean of the time series. This model is useful for such time series data, where volatility effects the dependent variables. In this work, we consider two variants of GARCH-M models: GARCH-ARIMA and GARCH-ARIMAX for forecasting of rice production.

\textbf{GARCH-ARIMA model}

The GARCH-ARIMA model is a non-linear hybrid model that incorporates an ARIMA model for modeling the mean equation along with the GARCH model for modeling the volatility of the time series. This model is particularly relevant for forecasting applications in economic and agricultural time series due to the presence of both autocorrelation and time varying volatility in the data.

\textbf{Mean Equation (ARIMA Component):}
\begin{equation}
    y_t = c + \sum_{i=1}^{p} \phi_i y_{t-i} + \sum_{j=1}^{q} \theta_j \epsilon_{t-j} + \epsilon_t,
\end{equation}
where $p$ and $q$ are the autoregressive and moving average orders, respectively.

\textbf{Variance Equation (GARCH Component):}

\begin{equation}
    \sigma_t^2 = \omega + \sum_{i=1}^{r} \alpha_i \epsilon_{t-i}^2 + \sum_{j=1}^{s} \beta_j \sigma_{t-j}^2.
    \label{eq:GARCH}
\end{equation}

\textbf{GARCH-ARIMAX Model}

The GARCH-ARIMAX model is an enhanced version GARCH-ARIMA by integrating exogenous variables that influence the dependent variable. This model is particularly useful when external factors contribute to variations in the series, such as rainfall in agricultural modeling. The GARCH-ARIMAX model consists of two primary components: the ARIMAX model for modeling the mean equation, and the GARCH model is used for modeling the conditional variance.

\textbf{Mean Equation (ARIMAX Component):}
\begin{equation}
    y_t = c + \sum_{i=1}^{p} \phi_i y_{t-i} + \sum_{j=1}^{q} \theta_j \epsilon_{t-j} + \sum_{k=1}^{K} \gamma_k X_{t,k} + \epsilon_t,
\end{equation}
where $X_{t,k}$ represents the exogenous variables.

Variance is modelled in an identical manner to the previous case (Eq~\ref{eq:GARCH}). For prediction of agricultural production, the GARCH-ARIMAX model helps us to measure the impact of rainfall variability on both the mean and volatility of crop yields, providing a better framework for risk assessment and decision-making.

\section{Data analysis}

In this study, we download monthly rainfall data (in millimeters) for Maharashtra spanning from 1962 to 2020 from the India Water Resources Information System (India-WRIS)(\url{https://indiawris.gov.in/wris/$\#$/timeseriesdata}). India-WRIS is an initiative taken by Government of India. They provide hydrological and meteorological data at varying spatial and temporal resolutions. Annual rice production data of Maharashtra for the same period is downloaded from the EPWRF India Time Series (https://epwrfits.in/Agriculture$\_$All$\_$India$\_$State$.$aspx). This website is maintained by the Economic and Political Weekly Research Foundation (EPWRF), and is well-known for long-term economic and agricultural statistics in India.

Prior to applying the econometric models, we examine the stationarity of the rice production time series data using the Augmented Dickey-Fuller (ADF) test \citep{dickey1979distribution}. As indicated from the test results suggest that the original series is non-stationary (\textit{ADF statistic} = -3.2824, \textit{p-value} = 0.08309). Non-stationarity implies that the mean and variance of the series do not remain constant over time, leading to spurious regression results when modeling its relationship with exogenous variables. To address this issue, we transform the series using first-order differencing $(\Delta Y_t = Y_t-Y_{t-1})$, which effectively removes trends and stabilizes the variance. The ADF test on the transformed series confirms stationarity (\textit{ADF statistic} = -5.9731, \textit{p-value} = 0.01). We ensure that our models yield accurate and significant results about the impact of rainfall variability on rice production.

The transformed series consists of 58 observations, which we divide into a training set of 55 data points, and a test set contains 3 data points for forecasting purposes. To capture rainfall variability, we construct four realized variance measures: $RV_i(t);\;i=1,2,3,4$ derived from the log-differenced monthly rainfall series. We employ four different time series models: ARIMA, ARIMAX, GARCH-ARIMA, and GARCH-ARIMAX, where ARIMA and GARCH-ARIMA serve as benchmarks, while ARIMAX and GARCH-ARIMAX incorporate exogenous realized variance measures to assess their predictive power. For the GARCH framework, we consider four variants—standard GARCH (sGARCH), exponential GARCH (eGARCH), threshold GARCH (gjrGARCH), and integrated GARCH (iGARCH)—to analyze different aspects of volatility dynamics. The models are estimated using the training data, and their in-sample fit is evaluated based on the Akaike Information Criterion (AIC). Forecasting performance is assessed using mean absolute error (MAE) for 1-step, 2-step, and 3-step ahead predictions on the test data. For ARIMAX and GARCH-ARIMAX, we incorporate the realized variance measures in two ways: using contemporaneous values corresponding to the differenced series and using lagged values to account for potential delayed effects of rainfall variability on rice production. This comprehensive approach allows us to evaluate the role of rainfall risk in predicting both the mean and volatility of agricultural output.

The first two columns of the table~\ref{tab:AICtable} present the AIC values for the simple econometric models (ARIMA and GARCH-ARIMA). ARIMA model represents the higher value of AIC indicating poorer fit compared to its alternatives. GARCH-ARIMA model. Integrating the rainfall risk measures into ARIMAX and GARCH-ARIMAX model produce mixed improvements in terms of AIC values. For all variants of GARCH models, realized volatility based on contemporary values yield superior results compared to the lagged values. Risk measures based on the first order differences of precipitation series demonstrate lower AIC values, establishing their efficiency in explaining rainfall effects. Exponential GARCH model demonstrates smaller values of AIC suggesting asymmetric effects of volatility explain fluctuations in rice productivity in a superior manner. Results corresponding to threshold GARCH and integrated GARCH models show that threshold effect and integrated persistence properties are not essential for this purpose.  

MAE values for the standard GARCH models are presented in tables~\ref{tab:MAEsGARCH1}--\ref{tab:MAEsGARCH2}. GARCH-ARIMA model generally with lower values of errors outperform the simple ARIMA model in many instances. Incorporation of risk measures do not always improve the prediction accuracy. $RV_3$ and $RV_4$ consistently yield lower MAE suggesting their better performance in capturing the effect of rainfall variability in agricultural production. Contemporary values of realized variances show better results compared to lagged values of the same suggesting immediate values is more impactful rather than delayed ones. Across all models, 1-step ahead forecast produce lowest MAE values. Increment of the error values to 2-step,3-step is sharper particularly for RV$_1$ and RV$_2$ explaining their relatively reduced variability for long-term prediction purposes. GARCH-ARIMAX models with contemporary RV$_3$ and RV$_4$ variables demonstrate most accurate forecast, explaining them as the most effective risk measure.

The MAE values for the exponential GARCH models are tabulated in tables~\ref{tab:MAEeGARCH1}--\ref{tab:MAEeGARCH2}. It is observed that integrating risk measures in simple models (ARIMA and GARCH-ARIMA) improve predictive accuracy, and the impact of incorporation measures varies based on which measure is used. Similar to the previous case, $RV_3$ and $RV_4$ generate lower MAE values specially for multi-step ahead forecasts in comparison to $RV_1$ and $RV_2$, explaining their importance of rainfall volatility on rice production. However, MAE values for GARCH-ARIMAX model with immediate $RV_4$ measures have the lowest MAE values for 1-step and 3-step ahead forecasting. Exponential GARCH model yield lower values of MAE compared to the standard GARCH model, specially for contemporary risk measures, explaining that integrating asymmetric volatility improves prediction accuracy.

The MAE values corresponding to the threshold GARCH models indicate that the predictive performance of integrating realized variances are effecting differently for different risk measures (see tables~\ref{tab:MAEgjrGARCH1}--\ref{tab:MAEgjrGARCH2}). Analogous to the previous cases, the measures $RV_3$ and $RV_4$ are outperforming $RV_1$ and $RV_2$ particularly for multi-step ahead forecasts. Simple models like ARIMA and GARCH-ARIMA outperform the econometric models with risk measures for certain instances in terms of MAE values. However, GARCH-ARIMAX model with contemporary $RV_4$ demonstrates substantial reduction in MAE for longer time horizons. Threshold GARCH model produces mixed results, as it outperforms the standard and exponential GARCH models in some cases. Overall, the result suggests that the threshold GARCH model is useful for modelling asymmetries in volatility and efficiency of the forecasting performance varies with the risk-measure and time horizon chosen.  

Tables~\ref{tab:MAEiGARCH1}--\ref{tab:MAEiGARCH2} revealed the MAE values for integrated GARCH model and it demonstrate the role of different risk measures for long-term prediction of rice production. Similar to the previous GARCH models, $RV_3$ and $RV_4$ outperform the remaining risk measures, indicating their efficiency in capturing the volatility structure of the variables. ARIMAX and GARCH-ARIMAX models with same-time realized variance values demonstrate improved accuracy compared to the simpler alternatives for long-term forecasting. For the shorter duration, ARIMA and GARCH-ARIMA models sometimes provide analogous or superior results indicating that the integrated GARCH model have more pronounced impact for longer time-horizons. Compared to previous GARCH models, this model depicts mixed performance, although it excels in some instances but not consistently outperform across all forecasting scenarios.

\section{Discussion}

In this study, we employ a wide variety of econometric models that incorporate rainfall risk measures to analyse the impact of rainfall variability on rice production in an Indian State, Maharashtra. Our contributions is this work are two fold. Firstly, we have four different econometric models and our result indicates the importance of integrating the risk measures into the models. Simple models like ARIMA and GARCH-ARIMA are treated as the baseline for prediction purposes, but their performance is weaker as the effect of rainfall risk measures is neglected. Forecasting accuracy of different econometric models are improved with the introduction of risk measures as exogenous variables in ARIMAX and GARCH-ARIMAX models. Realized risk measures based on the first order differences of rainfall values consistently outperform their alternatives in terms of lower AIC and reduced MAE values for long-term forecasts.  

Second contribution of our work is employing four different GARCH models to study how different types of volatility in rainfall series are impacting agricultural productivity. For example, asymmetric effects of volatility are incorporated in the model through the introduction of exponential GARCH model. Our analysis demonstrates that forecasting accuracy particularly for multi-step forecasts is improved for this model. The results indicate that eGARCH models explain fluctuations in rice production in a better manner compared to standard GARCH models. The result aligns with the intuition that extreme climatic events have different effect on agricultural productivity. The threshold GARCH model generates mixed results, indicating that the threshold effects in volatility are not crucial for improving predictive performance. Integrated GARCH model depicts the long-term persistence of volatility in the time series, and their predictive performance is enhanced for extended time horizons. Our results show that climate risks influence rice production beyond immediate seasonal effects, suggesting the requirement of policies that take long-term climatic variability into account.

From the policy perspective, our findings demonstrate the importance of using climate-aware forecasting models to support decision-making in agriculture. Traditional time series models that rely solely on historically available production data overlook the significant role of weather variability. By incorporating rainfall risk measures, policymakers and farmers can obtain more reliable predictions of rice production. Such policies are crucial for a state like Maharashtra, where monsoon rainfall is extremely important for agricultural productivity. This study provides robust evidence that incorporating rainfall risk measures into traditional econometric models improves the performance in terms of forecasting accuracy for prediction of rice production. Future research could be extension of this approach by exploring alternative formulations of risk measures, considering additional exogenous climatic variables, or employing analogous methodologies to other crops and regions.

\section*{Acknowledgements}
{This research is supported by DST-INSPIRE Fellowship Grant, Department of Science and Technology, Govt. of India (Grant no.: 190728) awarded to the first author.}
\section*{Declaration of generative AI and AI-assisted technologies in the writing process}
During the preparation of this work the author(s) used chatgpt(free version) in order to improve readability. After using this tool/service, the author(s) reviewed and edited the content as needed and take(s) full responsibility for the content of the publication.
\section*{Declaration of Interests:}
The authors declare no conflict of interest. The manuscript is not submitted or under review in any other journal simultaneously.
\section*{Author Contributions}
\textbf{Soham Ghosh:} Conceptualization, Data Curation, Formal analysis, Funding acquisition, Investigation, Methodology, Software, Validation, Visualization, Writing-Original draft preparation, Writing-Reviewing and Editing.
\par
\textbf{Sujay Mukhoti:} Conceptualization, Formal analysis, Investigation, Methodology, Project administration, Supervision, Validation, Writing-Original draft preparation, Writing-Reviewing and Editing.
\par
\textbf{Pritee Sharma:} Investigation, Project administration, Writing-Reviewing and Editing.
\section*{Data Availability:}
The datasets used in this study are publicly available. Further details on data processing and analysis are available from the corresponding author upon reasonable request.

\bibliographystyle{plainnat}
%\bibliography{References}

\newpage

\section{Appnedix : Tables}\label{Tables}

\begin{table}%[!ht]
\centering
\tiny
\caption{AIC values for different models.}
\begin{tabular}{|@{ }r@{ }||r@{ }|@{ }r@{ }||r@{ }|@{ }r@{ }||r@{ }|@{ }r@{ }||r@{ }|@{ }r@{ }||r@{ }|@{ }r@{ }||}
  \hline
   \multirow{2}{*}{AIC} & \multicolumn{2}{c||}{Simple Model} & \multicolumn{2}{c||}{RV$_1$} & \multicolumn{2}{c||}{RV$_2$} & \multicolumn{2}{c||}{RV$_3$} & \multicolumn{2}{c||}{RV$_4$} \\
   \cline{2-11}
   & ARIMA & GARCH-ARIMA & ARIMAx & GARCH-ARIMAx & ARIMAx & GARCH-ARIMAx & ARIMAx & GARCH-ARIMAx & ARIMAx & GARCH-ARIMAx \\ 
  \hline
  lag 1 (sGARCH) & 801.2619 & 14.7392 & 806.9639 & 14.7405 & 806.6786 & 14.7358 & 801.5409 & 14.6213 & 801.5720 & 14.6189 \\ \hline
  same time (sGARCH) & 801.2619 & 14.7392 & 795.5865 & 14.5290 & 795.1953 & 14.5231 & 801.0344 & 14.6484 & 801.0492 & 14.6489 \\ \hline
  lag 1 (eGARCH) & 801.2619 & 14.6482 & 806.9639 & 14.5377 & 806.6786 & 14.5215 & 801.5409 & 14.5072 & 801.5720 & 14.5212 \\ \hline
  same time (eGARCH) & 801.2619 & 14.6482 & 795.5865 & 14.5456 & 795.1953 & 14.4067 & 801.0344 & 14.6805 & 801.0492 & 14.6171 \\ \hline
  lag 1 (gjrGARCH) & 801.2619 & 14.6392 & 806.9639 & 14.7760 & 806.6786 & 14.7717 & 801.5409 & 14.6555 & 801.5720 & 14.6530 \\ \hline
  same time (gjrGARCH) & 801.2619 & 14.6392 & 795.5865 & 14.5630 & 795.1953 & 14.5570 & 801.0344 & 14.6828 & 801.0492 & 14.6833 \\ \hline
  lag 1 (iGARCH) & 801.2619 & 14.5699 & 806.9639 & 14.7068 & 806.6786 & 14.7020 & 801.5409 & 14.5893 & 801.5720 & 14.5870 \\ \hline
  same time (iGARCH) & 801.2619 & 14.5699 & 795.5865 & 14.4943 & 795.1953 & 14.4877 & 801.0344 & 14.6142 & 801.0492 & 14.6162 \\ \hline
\end{tabular}
\label{tab:AICtable}
\end{table}

\begin{table}%[!ht]
\centering
\tiny
\caption{MAE values for standard GARCH models with RV$_1$ and RV$_2$.}
\begin{tabular}{|@{ }r@{ }||r@{ }|r@{ }|r@{ }|@{ }r@{ }||r@{ }|r@{ }|r@{ }|@{ }r@{ }||}
  \hline
   \multirow{2}{*}{MAE} & \multicolumn{4}{c||}{RV$_1$} & \multicolumn{4}{c||}{RV$_2$}  \\
   \cline{2-9}
   & ARIMA & GARCH-ARIMA & ARIMAx & GARCH-ARIMAx & ARIMA & GARCH-ARIMA & ARIMAx & GARCH-ARIMAx \\ 
  \hline
  1 step(lag 1) & 114.2987 & 125.8558 & 152.0676 & 130.8162 & 270.6654 & 265.4435 & 232.8965 & 260.2492 \\ \hline
  2 step(lag 1) & 103.8309 & 162.0939 & 195.3841 & 166.7916 & 368.7445 & 408.7252 & 422.5288 & 408.7000 \\ \hline
  3 step(lag 1) & 114.0674 & 174.7338 & 147.6156 & 174.6215 & 467.5031 & 522.7875 & 475.8719 & 519.2039 \\ \hline
  1 step(same time) & 107.8242 & 97.0908 & 145.5931 & 102.1110 & 265.6363 & 312.5614 & 227.8673 & 307.2466 \\ \hline
  2 step(same time) & 121.1823 & 71.6080 & 212.7354 & 69.4247 & 465.3282 & 383.9998 & 519.1124 & 385.9012 \\ \hline
  3 step(same time) & 113.3607 & 50.7958 & 146.9089 & 47.7312 & 481.3305 & 392.5979 & 489.6994 & 392.3986 \\ \hline
\end{tabular}
\label{tab:MAEsGARCH1}
\end{table}

\begin{table}%[!ht]
\centering
\tiny
\caption{MAE values for standard GARCH models with RV$_3$ and RV$_4$.}
\begin{tabular}{|@{ }r@{ }||r@{ }|r@{ }|r@{ }|@{ }r@{ }||r@{ }|r@{ }|r@{ }|@{ }r@{ }||}
  \hline
   \multirow{2}{*}{MAE} & \multicolumn{4}{c||}{RV$_3$} & \multicolumn{4}{c||}{RV$_4$}  \\
   \cline{2-9}
   & ARIMA & GARCH-ARIMA & ARIMAx & GARCH-ARIMAx & ARIMA & GARCH-ARIMA & ARIMAx & GARCH-ARIMAx \\ 
  \hline
  1 step(lag 1) & 230.9177 & 228.7651 & 268.6866 & 255.4356 & 228.8440 & 225.9547 & 266.6130 & 251.5727 \\ \hline
  2 step(lag 1) & 130.4763 & 123.8605 & 191.9945 & 154.7327 & 128.7597 & 121.6780 & 191.6375 & 152.6944 \\ \hline
  3 step(lag 1) & 96.9015 & 91.9453 & 165.4009 & 113.9000 & 95.0856 & 89.7153 & 164.4914 & 112.1026 \\ \hline
  1 step(same time) & 231.2224 & 219.9568 & 268.9913 & 220.6701 & 229.8628 & 217.8995 & 267.6317 & 221.4061 \\ \hline
  2 step(same time) & 130.4871 & 124.2694 & 192.2885 & 124.1843 & 128.8001 & 122.0879 & 192.6158 & 123.5410 \\ \hline
  3 step(same time) & 96.9441 & 91.2089 & 165.6323 & 90.5604 & 95.2300 & 89.0809 & 165.2611 & 89.8045 \\ \hline
\end{tabular}
\label{tab:MAEsGARCH2}
\end{table}

\begin{table}%[!ht]
\centering
\tiny
\caption{MAE values for exponential GARCH models with RV$_1$ and RV$_2$.}
\begin{tabular}{|@{ }r@{ }||r@{ }|r@{ }|r@{ }|@{ }r@{ }||r@{ }|r@{ }|r@{ }|@{ }r@{ }||}
  \hline
   \multirow{2}{*}{MAE} & \multicolumn{4}{c||}{RV$_1$} & \multicolumn{4}{c||}{RV$_2$}  \\
   \cline{2-9}
   & ARIMA & GARCH-ARIMA & ARIMAx & GARCH-ARIMAx & ARIMA & GARCH-ARIMA & ARIMAx & GARCH-ARIMAx \\ 
  \hline
  1 step(lag 1) & 114.2987 & 125.8558 & 212.5434 & 194.2963 & 270.6654 & 265.4435 & 251.0317 & 237.3406 \\ \hline
  2 step(lag 1) & 103.8309 & 162.0939 & 139.7804 & 197.7361 & 368.7445 & 408.7252 & 358.8641 & 389.7212 \\ \hline
  3 step(lag 1) & 114.0674 & 174.7338 & 129.2517 & 178.1855 & 467.5031 & 522.7875 & 460.8737 & 522.4836 \\ \hline
  1 step(same time) & 107.8242 & 97.0908 & 202.0243 & 17.0230 & 265.6363 & 312.5614 & 201.6465 & 256.1439 \\ \hline
  2 step(same time) & 121.1823 & 71.6080 & 154.3295 & 138.2081 & 465.3282 & 383.9998 & 432.5051 & 408.4649 \\ \hline
  3 step(same time) & 113.3607 & 50.7958 & 126.1570 & 114.1596 & 481.3305 & 392.5979 & 458.8964 & 405.7767 \\ \hline
\end{tabular}
\label{tab:MAEeGARCH1}
\end{table}

\begin{table}%[!ht]
\centering
\tiny
\caption{MAE values for exponential GARCH models with RV$_3$ and RV$_4$.}
\begin{tabular}{|@{ }r@{ }||r@{ }|r@{ }|r@{ }|@{ }r@{ }||r@{ }|r@{ }|r@{ }|@{ }r@{ }||}
  \hline
   \multirow{2}{*}{MAE} & \multicolumn{4}{c||}{RV$_3$} & \multicolumn{4}{c||}{RV$_4$}  \\
   \cline{2-9}
   & ARIMA & GARCH-ARIMA & ARIMAx & GARCH-ARIMAx & ARIMA & GARCH-ARIMA & ARIMAx & GARCH-ARIMAx \\ 
  \hline
  1 step(lag 1) & 230.9177 & 228.7651 & 250.9196 & 238.2659 & 228.8440 & 225.9547 & 248.4906 & 277.4402 \\ \hline
  2 step(lag 1) & 130.4763 & 123.8605 & 140.5486 & 134.3390 & 128.7597 & 121.6780 & 138.6469 & 166.4490 \\ \hline
  3 step(lag 1) & 96.9015 & 91.9453 & 103.6639 & 99.5378 & 95.0856 & 89.7153 & 101.7197 & 122.8699 \\ \hline
  1 step(same time) & 231.2224 & 219.9568 & 250.9173 & 209.0021 & 229.8628 & 217.8995 & 249.5084 & 113.1772 \\ \hline
  2 step(same time) & 130.4871 & 124.2694 & 140.3995 & 117.9431 & 128.8001 & 122.0879 & 138.6835 & 58.3190 \\ \hline
  3 step(same time) & 96.9441 & 91.2089 & 103.5956 & 84.9633 & 95.2300 & 89.0809 & 101.8593 & 39.5813 \\ \hline
\end{tabular}
\label{tab:MAEeGARCH2}
\end{table}

\begin{table}%[!ht]
\centering
\tiny
\caption{MAE values for Threshold GARCH models with RV$_1$ and RV$_2$.}
\begin{tabular}{|@{ }r@{ }||r@{ }|r@{ }|r@{ }|@{ }r@{ }||r@{ }|r@{ }|r@{ }|@{ }r@{ }||}
  \hline
   \multirow{2}{*}{MAE} & \multicolumn{4}{c||}{RV$_1$} & \multicolumn{4}{c||}{RV$_2$}  \\
   \cline{2-9}
   & ARIMA & GARCH-ARIMA & ARIMAx & GARCH-ARIMAx & ARIMA & GARCH-ARIMA & ARIMAx & GARCH-ARIMAx \\ 
  \hline
  1 step(lag 1) & 114.2987 & 125.8558 & 196.3369 & 132.4557 & 270.6654 & 265.4435 & 188.6272 & 260.6364 \\ \hline
  2 step(lag 1) & 103.8309 & 162.0939 & 144.2896 & 167.3026 & 368.7445 & 408.7252 & 327.165 & 405.9633 \\ \hline
  3 step(lag 1) & 114.0674 & 174.7338 & 140.6663 & 175.0001 & 467.5031 & 522.7875 & 439.4098 & 521.7075 \\ \hline
  1 step(same time) & 107.8242 & 97.0908 & 189.8624 & 103.7653 & 265.6363 & 312.5614 & 183.5980 & 308.9754 \\ \hline
  2 step(same time) & 121.1823 & 71.6080 & 161.6410 & 70.3290 & 465.3282 & 383.9998 & 423.7486 & 382.1233 \\ \hline
  3 step(same time) & 113.3607 & 50.7958 & 139.9596 & 47.8548 & 481.3305 & 392.5979 & 453.2372 & 389.5243 \\ \hline
\end{tabular}
\label{tab:MAEgjrGARCH1}
\end{table}

\begin{table}%[!ht]
\centering
\tiny
\caption{MAE values for Threshold GARCH models with RV$_3$ and RV$_4$.}
\begin{tabular}{|@{ }r@{ }||r@{ }|r@{ }|r@{ }|@{ }r@{ }||r@{ }|r@{ }|r@{ }|@{ }r@{ }||}
  \hline
   \multirow{2}{*}{MAE} & \multicolumn{4}{c||}{RV$_3$} & \multicolumn{4}{c||}{RV$_4$}  \\
   \cline{2-9}
   & ARIMA & GARCH-ARIMA & ARIMAx & GARCH-ARIMAx & ARIMA & GARCH-ARIMA & ARIMAx & GARCH-ARIMAx \\ 
  \hline
  1 step(lag 1) & 230.9177 & 228.7651 & 312.9559 & 238.8071 & 228.8440 & 225.9547 & 228.6287 & 228.6197 \\ \hline
  2 step(lag 1) & 130.4763 & 123.8605 & 172.0559 & 146.7118 & 128.7597 & 121.6780 & 142.3848 & 142.3128 \\ \hline
  3 step(lag 1) & 96.9015 & 91.9453 & 124.9948 & 108.4830 & 95.0856 & 89.7153 & 105.2064 & 105.1164 \\ \hline
  1 step(same time) & 231.2224 & 219.9568 & 313.2606 & 223.5172 & 229.8628 & 217.8995 & 311.9010 & 222.8236 \\ \hline
  2 step(same time) & 130.4871 & 124.2694 & 172.0666 & 125.7498 & 128.8001 & 122.0879 & 170.3796 & 124.3127 \\ \hline
  3 step(same time) & 96.9441 & 91.2089 & 125.0374 & 91.6931 & 95.2300 & 89.0809 & 123.3233 & 90.2414 \\ \hline
\end{tabular}
\label{tab:MAEgjrGARCH2}
\end{table}

\begin{table}%[!ht]
\centering
\tiny
\caption{MAE values for integrated GARCH models with RV$_1$ and RV$_2$.}
\begin{tabular}{|@{ }r@{ }||r@{ }|r@{ }|r@{ }|@{ }r@{ }||r@{ }|r@{ }|r@{ }|@{ }r@{ }||}
  \hline
   \multirow{2}{*}{MAE} & \multicolumn{4}{c||}{RV$_1$} & \multicolumn{4}{c||}{RV$_2$}  \\
   \cline{2-9}
   & ARIMA & GARCH-ARIMA & ARIMAx & GARCH-ARIMAx & ARIMA & GARCH-ARIMA & ARIMAx & GARCH-ARIMAx \\ 
  \hline
  1 step(lag 1) & 114.2987 & 125.8558 & 198.5933 & 131.7870 & 270.6654 & 265.4435 & 186.3708 & 258.9003 \\ \hline
  2 step(lag 1) & 103.8309 & 162.0939 & 145.3977 & 169.4209 & 368.7445 & 408.7252 & 326.0167 & 410.9654 \\ \hline
  3 step(lag 1) & 114.0674 & 174.7338 & 141.3915 & 170.6596 & 467.5031 & 522.7875 & 438.6308 & 515.0209 \\ \hline
  1 step(same time) & 107.8242 & 97.0908 & 192.1188 & 101.1449 & 265.6363 & 312.5614 & 181.3416 & 311.9565 \\ \hline
  2 step(same time) & 121.1823 & 71.6080 & 162.7490 & 71.1615 & 465.3282 & 383.9998 & 422.6003 & 380.8716 \\ \hline
  3 step(same time) & 113.3607 & 50.7958 & 140.6848 & 49.1915 & 481.3305 & 392.5979 & 452.4582 & 389.3947 \\ \hline
\end{tabular}
\label{tab:MAEiGARCH1}
\end{table}

\begin{table}%[!ht]
\centering
\tiny
\caption{MAE values for integrated GARCH models with RV$_3$ and RV$_4$.}
\begin{tabular}{|@{ }r@{ }||r@{ }|r@{ }|r@{ }|@{ }r@{ }||r@{ }|r@{ }|r@{ }|@{ }r@{ }||}
  \hline
   \multirow{2}{*}{MAE} & \multicolumn{4}{c||}{RV$_3$} & \multicolumn{4}{c||}{RV$_4$}  \\
   \cline{2-9}
   & ARIMA & GARCH-ARIMA & ARIMAx & GARCH-ARIMAx & ARIMA & GARCH-ARIMA & ARIMAx & GARCH-ARIMAx \\ 
  \hline
  1 step(lag 1) & 230.9177 & 228.7651 & 315.2123 & 261.0867 & 228.844	& 225.9547 & 313.1387 & 256.7373 \\ \hline
  2 step(lag 1) & 130.4763 & 123.8605 & 173.2042 & 173.2150 & 128.7597 & 121.6780 & 157.1532 & 154.9512 \\ \hline
  3 step(lag 1) & 96.9015 & 91.9453 & 125.7738 & 115.5168 & 95.0856 & 89.7153 & 113.6070 & 123.9579 \\ \hline
  1 step(same time) & 231.2224 & 219.9568 & 315.5170 & 212.5725 & 229.8628 & 217.8995 & 314.1574 & 207.0190 \\ \hline
  2 step(same time) & 130.4871 & 124.2694 & 171.4876 & 119.8211 & 128.8001 & 122.0879 & 171.5280 & 115.6467 \\ \hline
  3 step(same time) & 96.9441 & 91.2089 & 125.8164 & 86.7086 & 95.2300 & 89.0809 & 124.1023 & 82.8698 \\ \hline
\end{tabular}
\label{tab:MAEiGARCH2}
\end{table}

\end{document}